# Initial characterization of interstellar comet 2I/Borisov

Piotr Guzik[1*], Michał Drahus[1*], Krzysztof Rusek[2], Wacław Waniak[1], Giacomo Cannizzaro[3,4], Inés Pastor-Marazuela[5,6]

[1] Astronomical Observatory, Jagiellonian University, Kraków, Poland

[2] AGH University of Science and Technology, Kraków, Poland

[3] SRON, Netherlands Institute for Space Research, Utrecht, the Netherlands

[4] Department of Astrophysics/IMAPP, Radboud University, Nijmegen, the Netherlands

[5] Anton Pannekoek Institute for Astronomy, University of Amsterdam, Amsterdam, the Netherlands

[6] ASTRON, Netherlands Institute for Radio Astronomy, Dwingeloo, the Netherlands

* These authors contributed equally to this work; email: piotr.guzik@doctoral.uj.edu.pl, drahus@oa.uj.edu.pl

**Interstellar comets penetrating through the Solar System had been anticipated for decades[1,2]. The discovery of asteroidal-looking 'Oumuamua[3,4] was thus a huge surprise and a puzzle. Furthermore, the physical properties of the 'first scout' turned out to be impossible to reconcile with Solar System objects[4–6], challenging our view of interstellar minor bodies[7,8]. Here, we report the identification and early characterization of a new interstellar object, which has an evidently cometary appearance. The body was discovered by Gennady Borisov on 30 August 2019 UT and subsequently identified as hyperbolic by our data mining code in publicly available astrometric data. The initial orbital solution implies a very high hyperbolic excess speed of ~32 km s$^{-1}$, consistent with 'Oumuamua[9] and theoretical predictions[2,7]. Images taken on 10 and 13 September 2019 UT with the William Herschel Telescope and Gemini North Telescope show an extended coma and a faint, broad tail. We measure a slightly reddish colour with a g′–r′ colour index of 0.66 ± 0.01 mag, compatible with Solar System comets. The observed morphology is also unremarkable and best explained by dust with a power-law size-distribution index of –3.7 ± 1.8 and a low ejection speed (44 ± 14 m s$^{-1}$ for $\beta$ = 1 particles, where $\beta$ is the ratio of the solar gravitational attraction to the solar radiation pressure). The nucleus is probably ~1 km in radius, again a common value among Solar System comets, and has a negligible chance of experiencing rotational disruption. Based on these early characteristics, and putting its hyperbolic orbit aside, 2I/Borisov appears indistinguishable from the native Solar System comets.**

On 8 September 2019 at 04:15 UT, we were alerted by our software Interstellar Crusher (see Methods) of a possible new hyperbolic object gb00234. Within less than 4 d, the orbit became reliable enough to trigger the first announcements[10,11], and subsequently, the body received an official name 2I/Borisov. As of 20 September 2019 at 12:00 UT, 447 published astrometric positions collected over a 21.1 d interval[11–15] are demonstrably incompatible with a parabolic orbit (Fig. 1). In the absence of non-gravitational forces, the residuals show a very strong systematic trend and reach up to 20 arcsec. Inclusion of non-gravitational forces greatly reduces the residuals, but a small systematic trend is still present. More importantly, the resulting non-gravitational accelerations $A_1 = 7.43 \pm 0.07 \times 10^{-4}$ au d$^{-2}$ and $A_2 = -1.83 \pm 0.01 \times 10^{-4}$ au d$^{-2}$ given at 1 au from the Sun (see Methods) are implausibly high, exceeding the largest measured non-gravitational accelerations of comets[16] by two to three orders of magnitude and comparable to the sunward gravitational acceleration at 1 au. However, the data are accurately fitted by an unrestricted, purely gravitational solution (Fig. 1) that implies a strongly hyperbolic orbit with an eccentricity of $3.38 \pm 0.02$ (Table 1). This strong hyperbolicity cannot be attributed to gravitational perturbations from the Solar System's planets because the body travels from a direction far from the ecliptic plane. Thus, the only viable explanation is the arrival from outside the Solar System. The huge eccentricity together with a moderate perihelion distance of $2.012 \pm 0.004$ au (Table 1) imply a hyperbolic excess speed of ~32 km s$^{-1}$. The body entered the Solar System from a direction ~75° away from the Solar apex with the asymptotic radiant at J2000.0 right ascension (RA) = 02 h 12 m and declination (dec) = 59.4° in the constellation of Cassiopeia. For the most up-to-date orbital parameters, readers are referred to the online databases of the Minor Planet Center or Jet Propulsion Laboratory.

**Table 1 | Hyperbolic heliocentric orbital elements of 2I/Borisov calculated for the osculation epoch 2019 September 20.0 TT**

| | |
|---|---|
| $T$ | 2019 Dec. 8.42 ± 0.11 TT |
| $e$ | 3.3790 ± 0.020 |
| $q$ | 2.0119 ± 0.0044 au |
| $\omega$ | 209.001 ± 0.100° (J2000.0) |
| $\Omega$ | 308.195 ± 0.040° (J2000.0) |
| $i$ | 44.004 ± 0.041° (J2000.0) |

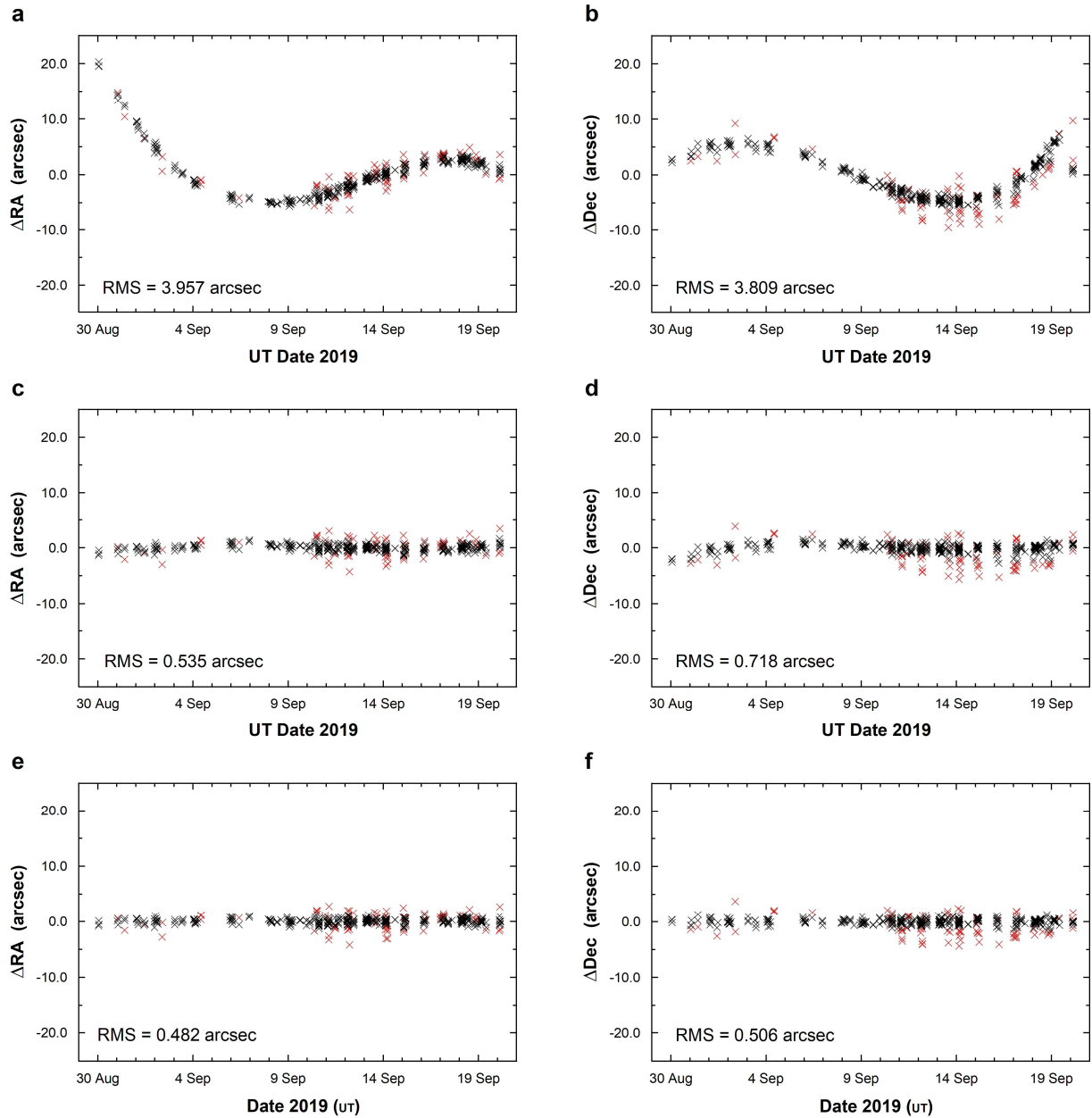

**Fig. 1 | Astrometric residuals of 2I/Borisov calculated for three different orbital solutions. a,b**, The residuals calculated for a parabolic solution without non-gravitational forces. **c,d**, The residuals for a parabolic solution with non-gravitational forces. **e,f**, The residuals for an unrestricted solution without non-gravitational forces. Residuals are presented separately in RA (**a,c,e**) and dec (**b,d,f**). Black symbols denote data used in the computation and red symbols denote rejected outliers. r.m.s., root mean squared.

We observed this object using the 4.2 m William Herschel Telescope (WHT) on La Palma and the 8.2 m Gemini North Telescope at Maunakea in the Sloan Digital Sky Survey (SDSS) g′ and r′ bands[17]. WHT data were obtained with the Auxiliary-port CAMera (ACAM) on 10 September 2019 at 05:38 UT and on 13 September 2019 at 05:47 UT (observation midpoints). The first set comprises ten sidereal-tracked 60 s exposures, of which five were obtained in the g′ band and five in the r′ band, whereas the second set contains 40 sidereal-tracked 20 s exposures, 20 obtained in g′ and 20 in r′. Gemini data were collected with the Gemini Multi-Object Spectrograph (GMOS-N) on 10 September 2019 at 14:57 UT (midpoint) with non-sidereal tracking and comprise four g′-band and four r′-band exposures taken with 60 s integration time. The datasets were obtained at low elevation (22° to 31°) in morning twilight (solar elevation from −19° to −12°). At the time of the observations, the helio- and geocentric distances of 2I/Borisov were equal to 2.8 and 3.4 au, respectively, the phase angle was ~15° and the apparent motion was ~75 arcsec h$^{-1}$. The images were corrected for overscan, bias and flatfield in the standard fashion, and then a global background level was subtracted from each frame.

In Fig. 2 we show median-stacked g′-band and r′-band images from Gemini, which have a better signal-to-noise ratio than the WHT images. The latter are presented in Supplementary Fig. 1. The images reveal an extended coma and a broad, short tail emanating in roughly antisolar direction. We see no clear difference in morphology in the two bands. The comet was measured photometrically in each individual exposure from the two telescopes and two nights. We consistently used a 5,000 km (~2 arcsec) radius photometric aperture and determined the brightness against background stars available in the SDSS photometric catalogue[18]. As a result, we obtained dataset-averaged AB magnitudes g′ = 19.38 ± 0.01 and r′ = 18.71 ± 0.01 for the WHT observations on 10 September 2019 at 05:38 UT, g′ = 19.38 ± 0.01 and r′ = 18.72 ± 0.01 for the Gemini observations on 10 September 2019 at 14:57 UT, and g′ = 19.32 ± 0.02 and r′ = 18.67 ± 0.01 for the WHT observations on 13 September 2019 at 05:47 UT. Photometric magnitudes can be used to estimate the nucleus size. Following two independent approaches, we have found that the nucleus of 2I/Borisov is most likely ~1 km in radius (see Methods). However, it should be noted that these approaches are inherently very uncertain. Our photometric measurements give a consistent g′–r′ colour index, with an average value of 0.66 ± 0.01 mag. The colour is slightly redder than the solar g′–r′ = 0.45 ± 0.02 mag[19] and implies a positive spectral slope $S'$ ~ 12.5% per 100 nm (see

Methods), in good agreement with an independent spectroscopic determination[20]. Within the errors, the same colour is obtained for other photometric apertures as well (we made measurements for the apertures ranging from 3,000 to 15,000 km in radius), implying a uniform colour of the coma. This, and the similarity of morphology in both bands, are both indicative of dust-dominated activity in our data. Monte Carlo modelling of the object's dust environment with our established code[21] has revealed a power-law particle size-distribution exponent of $-3.7 \pm 1.8$ and a $44 \pm 14$ m s$^{-1}$ ejection speed applicable to $\beta = 1$ particles (where $\beta$ is the ratio of the solar gravitational attraction to the solar radiation pressure; see Methods).

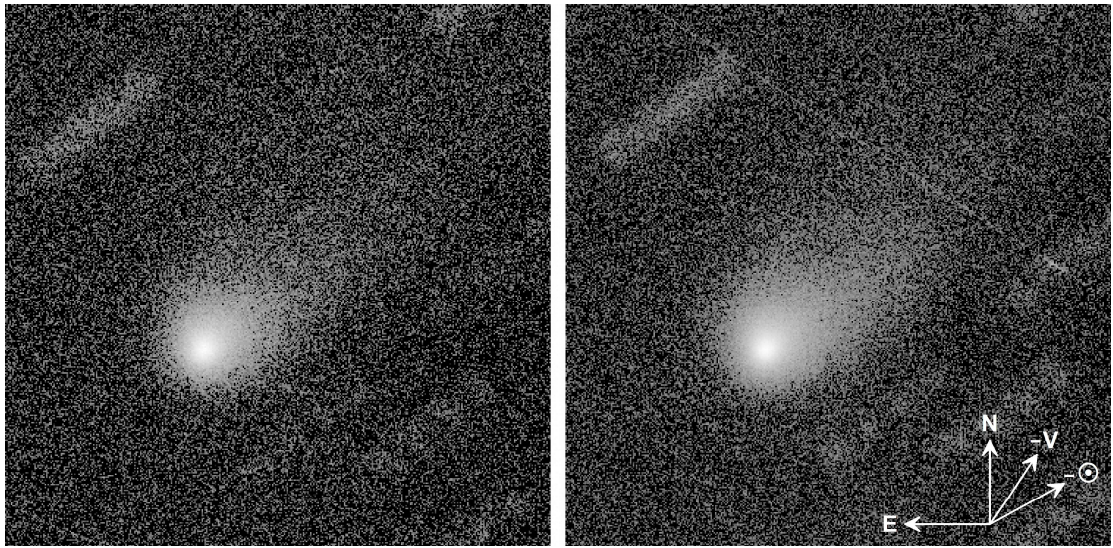

**Fig. 2 | Median-stacked images of 2I/Borisov from Gemini North.** The left panel shows the image in the g′ band and the right panel shows the image in the r′ band. Both panels subtend 1.0 × 1.0 arcmin and were scaled logarithmically. Arrows show the directions of north (N) and east (E), the projected antisolar vector (–☉) and the negative of the orbital velocity vector (–V). The pixel scale is 0.16 arcsec px$^{-1}$ and the seeing was 1.8 arcsec.

The dynamical properties and morphology of 2I/Borisov make it clear that the body is the first certain case of an interstellar comet, and the second known interstellar minor body identified in the Solar System (after 'Oumuamua[3,4]). Evidently, the extended coma and the broad tail stand in stark contrast with the purely asteroidal appearance of 'Oumuamua. The estimated nucleus size is common among Solar System's comets[22,23]. Adopting the formalism of rotational disruption probability[24] and assuming typical properties of Solar System comets, we have estimated the probability of rotational disruption during the Solar System flyby to be smaller than 1% (see Methods). The measured colour is consistent with the colours of the Solar System's comets[25–27] and falls only slightly redwards of the median and average values of the observed g′–r′ distribution (Fig. 3). The same similarity can be noticed for the dust coma parameters[28,29]. These facts are remarkable in and of themselves, and especially remarkable after 'Oumuamua, the multiple peculiarities of which[4–6] prompted us to rethink our entire view of the nature of interstellar interlopers[7]. However, 2I/Borisov appears completely similar to the native Solar System's comets.

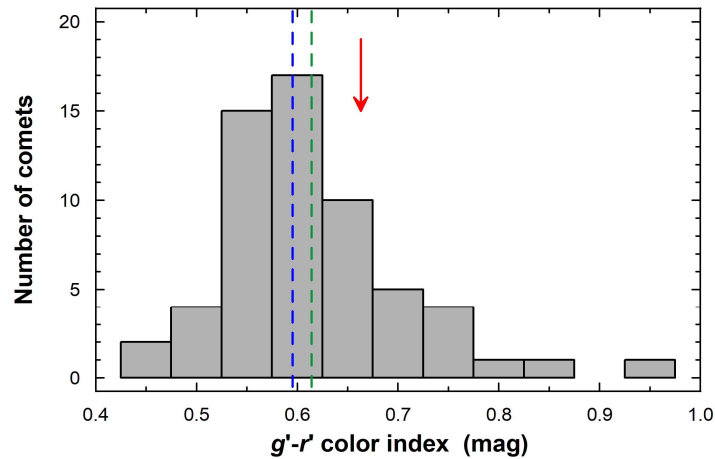

**Fig. 3 | Colour of 2I/Borisov in the context of Solar System comets.** The histogram shows the g′–r′ colour index distribution for 60 comets[25–27] (long-period, Jupiter-family and active Centaurs). It was calculated from the colour indices reported in the Johnson–Cousins ($UBVR_CI_C$) and the original SDSS (ugriz) photometric systems using standard transformation formulae[36,45]. Whenever available, multiple measurements of a single object were filtered and combined to minimize the colour uncertainty. The sample has an r.m.s. error of g′–r′ equal to 0.036 mag and the error does not exceed 0.075 mag for any object (the latter criterion resulted in the rejection of three comets). Dashed lines show the median (blue) and average (green) values of the distribution, and the red arrow indicates the measured g′–r′ colour index of 2I/Borisov.

Comet 2I/Borisov was discovered on its way to perihelion (8 December 2019 UT at 2.0 au) and before the closest approach to Earth (28 December 2019 UT at 1.9 au); thus, the overall visibility will be gradually improving. The body is destined for an intensive observing campaign lasting many months (Supplementary Fig. 2) that will allow us to gain groundbreaking insights into the physical properties of interstellar comets and exosolar planetary systems in general. This discovery is in line with the detection statistic of one interstellar object per year proposed by several authors[30,31] after the discovery of 'Oumuamua, which is an order of magnitude higher than the most optimistic pre-'Oumuamua estimates[32]. It also shows that cometary nature of these bodies might be a common characteristic, in agreement with the original expectations[1,2]. More discoveries are expected in the near future thanks to the Large Synoptic Survey Telescope.

## Methods

**Interstellar Crusher**

Interstellar Crusher is a custom Python3 code running on Windows Subsystem for Linux. It continuously monitors the Possible Comet Confirmation Page and computes orbits of newly discovered minor bodies using Bill Gray's Find_Orb (https://github.com/Bill-Gray/find_orb/commit/abe3f5847aad4c39f1d82239bf343c876b3d1bd0). Detection of a possible interstellar object triggers an alarm that is sent via e-mail.

**Orbit**

We computed the orbit using Bill Gray's Find_Orb with planetary perturbations. The initial input dataset comprised 447 astrometric positions obtained at 45 different stations that were publicly available as of 20 September 2019 at 12:00 UT. We computed parabolic and unrestricted solutions without non-gravitational forces and a parabolic solution with non-gravitational forces defined according to the standard Marsden–Sekanina model[33]. Given the absence of reliable uncertainties of individual observations, we assumed equal weights. Astrometric positions with residuals greater than 1.5 arcsec with respect to the unrestricted solution without non-gravitational forces were iteratively rejected, resulting in a restricted dataset of 362 measurements. This dataset was used for the final computations, the results of which are presented in Fig. 1 and Table 1. The reported uncertainties of the orbital elements were estimated by the least squares method from the astrometric residuals.

**Photometry and colour**

Individual calibrated exposures were first scrutinized for background stars, cosmic-ray hits and other artefacts in the comet's photometric aperture. As a result of this procedure, we rejected one g′-band frame from the Gemini dataset, one g′-band and one r′-band frame from the first WHT dataset, and accepted all frames from the second WHT dataset. For each dataset, we identified a set of field stars available in the SDSS photometric catalogue[18] that were brighter than the comet and had a g′–r′ colour index between 0.3 and 1.0 mag. We identified 7 and 9 suitable stars for the first and second WHT dataset, respectively, and 4 stars for the Gemini dataset. Photometric measurements of the comet were done using a circular aperture with a 5,000 km (~2.0 arcsec) radius. The stars were measured in a larger, 4.0 arcsec radius aperture that contained >99% of the flux (determined from the curve of growth). Sky background level was estimated (as a mean value) in an annular aperture with a radius of 7.0–9.0 arcsec for the comet and 5.0–8.0 arcsec for the stars. We masked the regions of the background aperture contaminated by the comet's tail, faint field stars and cosmic-ray hits. Centroids were measured from a 1.0 arcsec radius aperture. Differential magnitudes of the comet were calculated for each individual image and then dataset averaged (with equal weights) and compared with the SDSS catalogue magnitudes of the reference stars. The photometric uncertainties were propagated from the individual differential measurements and SDSS uncertainties of the reference stars.

Colour index ($m_1 - m_2$) can be easily converted to the normalized reflectance slope $S'$. By definition

$$(m_1 - m_2) - (m_1 - m_2)_\odot = -2.5 \log(S_1/S_2)$$

where $(m_1 - m_2)_\odot$ is the solar colour index, $S_\lambda = 1 + S'(\lambda - \lambda_0)/\Delta\lambda_0$ is the normalized reflectance at wavelength $\lambda$, $\Delta\lambda_0$ defines the standard wavelength interval of $S'$ and $\lambda_0$ is the normalization wavelength. Using the customary values $\lambda_0 = 550$ nm and $\Delta\lambda_0 = 100$ nm and substituting the measured g′–r′ colour index of comet 2I/Borisov equal to 0.66 mag along with the effective filter wavelengths $\lambda_{g'} = 475$ nm and $\lambda_{r'} = 630$ nm and the solar g′–r′ = 0.45 mag[19], we calculated the corresponding slope $S' = 12.5\%$ per 100 nm.

**Nucleus size estimation**

To estimate the size of the nucleus, we followed a simple approach[2] that connects theoretical sublimation rate of water from a unit surface area[34] with an empirical correlation of the observed total visual magnitudes with the total water production rates[35]. By solving the standard energy budget equation[34] with an assumed Bond albedo of 4% and emissivity of 100%, we find that the average water sublimation flux from a spherical non-rotating nucleus is $1.86 \times 10^{26}$ molecules s$^{-1}$ km$^{-2}$ at the heliocentric distance applicable to our observations. From our Gemini data, we estimate the total visual magnitude to be ~17.5, which is the asymptotic magnitude from the curve of growth transformed from the SDSS g′ and r′ bands to the Johnson V band[36]. Taking into account the geo- and heliocentric distances at the time of the observation, this magnitude corresponds to the water production rate of ~$10^{27}$ molecules s$^{-1}$ according to an empirical relation[35]. By comparing the theoretical and observed water production rates, we estimate the area of the sublimating surface to be ~5 km$^2$, which corresponds to ~1 km radius nucleus with 30% active fraction, assuming negligible contribution of icy grains in the coma.

The same result is obtained from a comparison of 2I/Borisov with the well-studied comet Hale–Bopp in terms of the Afρ parameter[37]. From our r′-band magnitudes measured in the 5,000 km radius aperture, we calculate Afρ ~ 100 cm, and similar for other aperture sizes. The Afρ of comet Hale–Bopp was ~10,000 cm at the heliocentric distance and phase angle compatible with our observations[38]. Simple scaling of the ~30 km in radius nucleus of the latter comet[39] suggests the nucleus radius of 2I/Borisov of ~1 km.

**Probability of rotational disruption**

Rotational disruption occurs when the nucleus rotation rate becomes too high for the self-gravity and tensile strength to keep the body intact[40]. The probability of disruption P has been formulated[24] as the ratio of the expected change in the rotation frequency $\Delta\omega$ to the total extent of tolerable frequency regime, limited by the negative and positive critical frequency $\omega_{\text{crit}}$. Thus

$$P = \frac{\Delta\omega}{2\omega_{\text{crit}}}$$

The two components of this equation can be calculated in a relatively straightforward manner using standard formulae[5,41–44]. Substituting in these formulas the estimated nucleus radius of ~1 km and adopting typical properties of Solar System comets[24], we find the probability of rotational disruption of 2I/Borisov to be <1%.

**Dust modelling**

We modelled the dust environment of 2I/Borisov using a Monte Carlo approach[21]. The simulations were done with $2 \times 10^6$ power-law distributed dust particles spanning a size range $a$ from 0.5 μm to 1 mm. We assumed the density of the dust material to be 1,000 kg m$^{-3}$, the scattering efficiency for radiation pressure to be 1.0 and a size-dependent dust ejection speed (at the boundary of the collisional zone) $v_e \sim a^{-0.5}$. The dust emission rate is assumed to be inversely proportional to the square of the heliocentric distance with the earliest particles ejected 70 d before the observation. Under these assumptions, we investigated three different ejection patterns: (1) isotropic; (2) into the subsolar hemisphere with the emission rate proportional to the cosine of the solar zenith distance; and (3) into a sunward conical jet with an opening angle of 30° and a constant intensity. The model was fitted to our highest signal-to-noise ratio r′-band image created from the Gemini data (Fig. 2). The best fit was obtained for the hemispheric pattern (Supplementary Fig. 3), though the fit of the isotropic emission is nearly as good. As a result of this procedure, we retrieved a power-law particle size-distribution exponent of $-3.7 \pm 1.8$ and the ejection speed for particles of a given size, equal to $44 \pm 14$ m s$^{-1}$ for $\beta = 1$ ($a = 1.2$ μm).

## Acknowledgements

Based in part on observations obtained at the Gemini Observatory, which is operated by the Association of Universities for Research in Astronomy, Inc., under a cooperative agreement with the NSF on behalf of the Gemini partnership: the National Science Foundation (United States), the National Research Council (Canada), CONICYT (Chile), Ministerio de Ciencia, Tecnología e Innovación Productiva (Argentina), and Ministério da Ciência, Tecnologia e Inovação (Brazil). The William Herschel Telescope is operated on the island of La Palma by the Isaac Newton Group of Telescopes in the Spanish Observatorio del Roque de los Muchachos of the Instituto de Astrofísica de Canarias. We thank J. Blakeslee for rapid evaluation and approval of our Gemini North director's discretionary time request and P. Jonker for sharing time on the William Herschel Telescope. We also thank the staff of both observatories for assistance and vital contributions to making these observations possible. M.D. and P.G. are grateful for support from the National Science Centre of Poland through SONATA BIS grant no. 2016/22/E/ST9/00109 and Polish Ministry of Science and Higher Education grant no. DIR/WK/2018/12. G.C. acknowledges support from European Research Council Consolidator Grant 647208. I.P.-M. acknowledges funding from the Netherlands Research School for Astronomy (grant no. NOVA5-NW3-10.3.5.14).

## Author contributions

## Data availability

The ACAM data are available from the corresponding authors upon reasonable request. The GMOS-N raw data will be available in the Gemini Observatory archive at https://archive.gemini.edu after the expiration of the 12 month proprietary period.

# Supplementary Information

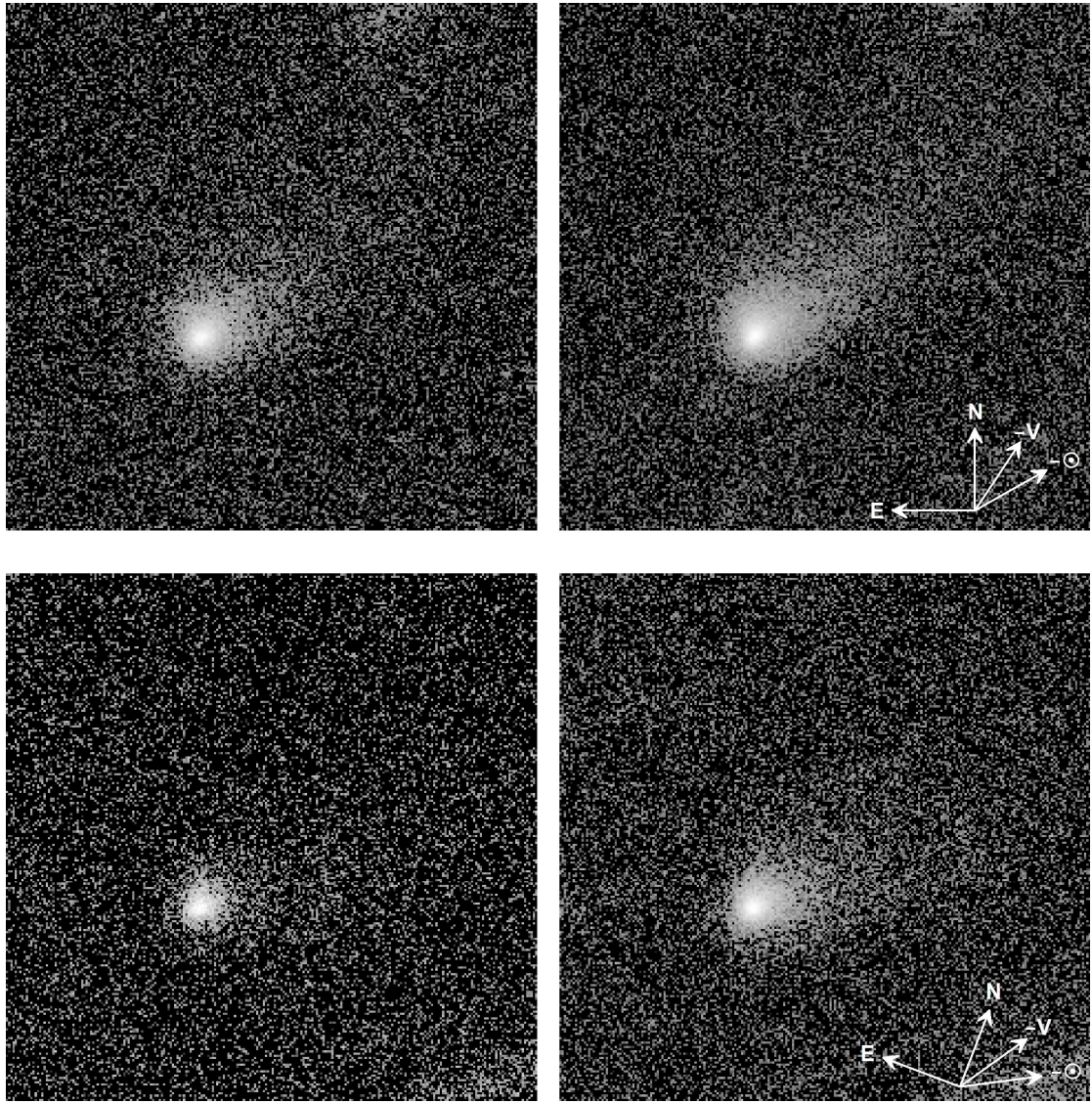

**Supplementary Fig. 1 | Median-stacked images of 2I/Borisov from the William Herschel Telescope.** The top panels show the images obtained on 10 September 2019 at 05:38 UT and the bottom panels show the images obtained on 13 September 2019 at 05:47 UT (observation mid-points). The images in the left panels were obtained in the g′ band and the images in the right panels were obtained in the r′ band. All panels subtend 1.0 × 1.0 arcmin and are scaled logarithmically. Arrows show the directions of north (N) and east (E), the projected antisolar vector (–☉) and the negative of the orbital velocity vector (–V). The pixel scale is 0.25 arcsec px$^{-1}$ and the seeing was 1.3 arcsec on both nights. The data from 13 September 2019 UT are affected by stronger twilight and moonlight, and possibly also by a thin cirrus cloud, resulting in the reduced apparent coma size.

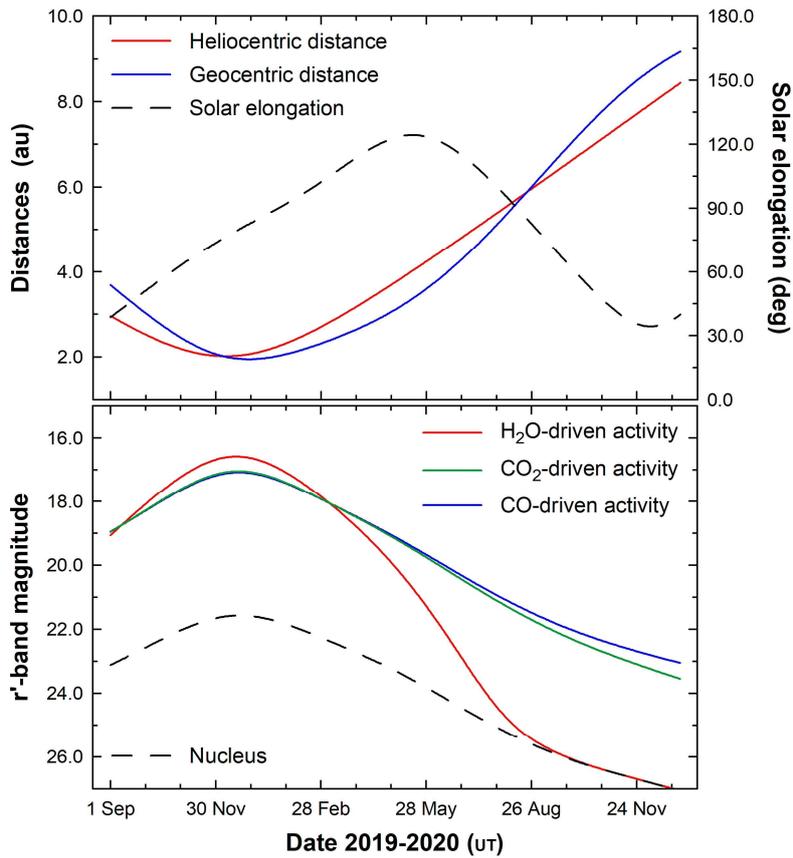

**Supplementary Fig. 2 | Visibility prospects of 2I/Borisov for years 2019 – 2020.** Top panel presents basic geometric circumstances and bottom panel shows r′-band magnitude in a 5,000-km radius aperture extrapolated from our Gemini North and WHT photometric data of 10 September 2019 UT. The magnitude is forecasted using three different activity models, in which the heliocentric brightness of the coma follows the sublimation curves of $H_2O$, $CO_2$ and CO computed in a standard manner with an assumed Bond albedo of 4% and emissivity of 100%. The actual observed magnitude additionally accounts for the nucleus brightness (we assumed a 4% geometric albedo and a 0.04 mag deg$^{-1}$ phase function), phase function of the coma (0.02 mag deg$^{-1}$), and geocentric term (inverse square law for brightness).

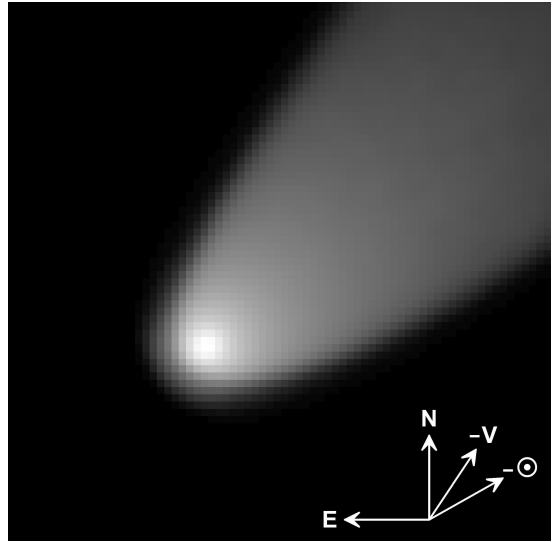

**Supplementary Fig. 3 | Synthetic image of 2I/Borisov computed with our Monte-Carlo code.** The model was fitted to the r′-band image obtained at Gemini North on 10 September 2019 at 14:57 UT (Fig. 2, right panel) and has the same orientation and scale. See Methods for details of the computation.